\documentclass{aa}  

\usepackage{graphicx}
\usepackage{txfonts}
\usepackage{float}
\usepackage{mathrsfs,amsmath}
\usepackage{upgreek}
\usepackage{float}
\usepackage{xcolor}
\usepackage[utf8]{inputenc}
\usepackage[colorlinks=true, citecolor=blue]{hyperref}
\hypersetup{colorlinks=true,linkcolor=blue,urlcolor=blue}
\begin{document}

    \title{\texttt{gateau}: an observation simulator for ground-based submillimeter astronomy with integral field units and kinetic inductance detectors}
    \titlerunning{\texttt{gateau}: GPU-accelerated time-dependent observation simulator}

    \author{A. Moerman\fnmsep\thanks{Corresponding author \email{A.Moerman-1@tudelft.nl}}
          \inst{1}
          \and
          N. Soshnin\inst{1}
          \and
          S. A. Brackenhoff\inst{1}
          \and
          S. O. Dabironezare\inst{1,2}
          \and
          K. Karatsu\inst{2}
          \and
          L. H. Marting\inst{1,2}
          \and
          S. A. H. de Rooij\inst{3,2,1}
          \and
          M. Roos\inst{1,4}
          \and
          B. R. Brandl\inst{5,6}
          \and
          A. Endo\inst{1}
          }

    \institute{Faculty of Electrical Engineering, Mathematics and Computer Science, Delft     University of Technology, Mekelweg 4, 2628 CD Delft, The Netherlands
         \and    
    SRON—Netherlands Institute for Space Research, Niels Bohrweg 4, 2333 CA Leiden, The Netherlands
        \and
    Huygens-Kamerlingh Onnes Laboratory, Leiden Institute of Physics, Leiden University, PO Box 9504, 2300 RA Leiden, The Netherlands
        \and
    Faculty of Applied Sciences, Delft University of Technology, Lorentzweg 1, 2628 CJ Delft, The Netherlands
        \and
    Leiden Observatory, Leiden University, PO Box 9513, 2300 RA Leiden, The Netherlands
        \and
    Faculty of Aerospace Engineering, Delft University of Technology, Kluyverweg 1, 2629 HS Delft, The Netherlands\\
    }

   \date{Received; accepted}

  \abstract
   {Submillimeter (submm) integral field units (IFUs) utilising kinetic inductance detectors (KIDs) are a promising instrument architecture for the study of galaxies, galaxy clusters, and the large-scale structure of the Universe. In order to design successful experiments targeting these science cases, several aspects such as instrument design, observation and calibration strategies, and data reduction pipelines must be collectively developed, tested, and optimised. This can be achieved through end-to-end simulations of the experiment, allowing for quantitative assessment of the aforementioned aspects.}
   {To this end, we have developed \texttt{gateau}, a modular, flexible, and efficient simulator for submm IFU observations of astronomical sources. The simulator consists of a Python interface, powered by a C/C++ backend that uses CUDA for GPU-acceleration, and is publicly available and fully open-source.}
   {\texttt{gateau} simulates observations by taking user input such as an astronomical source, a set of atmospheric screens, a scan pattern, and telescope and instrument parameters. The source signal is propagated through a dynamical model of the atmosphere and optical path at the telescope using a customised radiative transfer cascade. A dispersive element model, which can be a filterbank, grating, or user-supplied model, is used to calculate the total power per spectral channel, for each spatial pixel on the IFU. A physically motivated photon-noise model is used to add a white noise component to the received power. Detector noise is added as temporally correlated pink noise. The output is stored in the form of time-ordered datasets.}
   {We validated \texttt{gateau} against observations with DESHIMA 2.0, a superconducting, ultra-wideband spectrometer utilising KIDs and on-chip filterbank technology. We show that we can reproduce real observations of the atmosphere and Uranus with \texttt{gateau} simulations. Lastly, we present a use case to show how \texttt{gateau} can simulate long observations in timespans orders of magnitude smaller than the observation time itself, highlighting its applicability and efficiency.}
   {}


   \maketitle
%

\section{Introduction}
Over the last decade, kinetic inductance detectors (KIDs)~\citep{Day2003} have shown to be a promising technology for observing the submillimeter (submm) sky, offering extremely high sensitivity~\citep{Baselmans2022} and the ability to operate using a highly multiplexed readout allowing for large detector arrays~\citep{vanRantwijk2016,Baselmans2017}. Instruments such as NIKA2~\citep{Calvo2016} and Z-spec~\citep{Inami2008} have obtained science-grade observational data and others such as CONCERTO~\citep{Ade20}, AMKID~\citep{Reyes2026}, and DESHIMA 2.0~\citep{Taniguchi_2022,Karatsu26} have seen first light (see for example~\cite{Hu2024,Dsert2025} for CONCERTO and~\cite{Moerman2025} for DESHIMA 2.0) and are set to produce science-grade data soon. Moreover, recent advances in on-chip filterbank designs~\citep{Marting2024}, fabrication techniques~\citep{Scholtenhuis26}, and optical design approaches~\citep{Dabironezare2026} facilitate the development of ultra-wideband on-chip submm integral field units (IFUs) using KID technology, which are crucial for the goal of large 3-dimensional sky surveys in the submm range~\citep{Jovanovic2023}.

For ground-based submm IFU observations, the atmosphere is a major and difficult obstacle which needs to be corrected for. The effect of the atmosphere is twofold: attenuation of the astronomical source signal and addition of parasitic signal, several orders of magnitude larger than the astronomical signal in most cases, through the emissivity of the atmosphere. These two effects are time-dependent, causing the received signal to fluctuate in time. Additionally, multiple optical stages are usually encountered after entering the telescope and also cause the parasitic emission from the environment to mix with the signal from the sky. Examples of this are contamination by thermal emission through spillover losses on optical elements and reflectors, Ohmic losses on reflector surfaces, and dissipative losses inside lenses or cryostat windows. 

Furthermore, the quantum nature of the electromagnetic field causes photon noise~\citep{Yates2011}. The KIDs themselves also generate noise, such as quasiparticle generation-recombination noise~\citep{deVisser2012} and temporally correlated two-level system (TLS) noise~\citep{Gao2008-qh,Sueno2022} arising from quantum state fluctuations in the dielectric material of the detector. These noise sources further contaminate the sky signal and make retrieval of the astronomical source signal a non-trivial task, requiring specialised reduction and analysis pipelines before scientific interpretation can take place.

It is crucial to test these reduction and analysis pipelines on synthetic data, which requires an observation simulator that can produce time-ordered datasets (TODs) for each spectral channel (voxel) per spatial pixel (spaxel) on the IFU, resembling the actual output TODs of the real instrument as close as possible. The key advantage compared to testing on observed data is the ability to know the ground truth underlying the data produced by the simulator. This allows for quantitative assessment of the aforementioned reduction and analysis pipelines, but also instrument design, such as filterbank parameters, and observation strategies for particular astronomical sources, to name a few.

Observation simulators exist for a wide range of instruments, such as TOAST~\citep{TOAST}, TiEMPO~\citep{Huijten2022}, KSIM~\citep{Hofmann2023}, the MOMOS simulator~\citep{Kim25}, \texttt{MIRISIM}~\citep{Klaassen2020}, SOPHISM~\citep{BlancoRodrguez2018}, and \texttt{maria}~\citep{vanMarrewijk2024}, to name a few. Of these, only TiEMPO, KSIM, and the MOMOS simulator are specifically designed for instruments containing KIDs. KSIM, originally developed as a simulator package for the KIDSpec~\citep{OBrien2020} instrument, and the MOMOS simulator both focus on echelle spectrographs with optical/near-infrared KID arrays, rendering them unsuitable for simulating submm IFUs with filterbank technology. TiEMPO was designed to simulate single-pixel submm spectrometers using filterbank technology and KIDs, but its performance was never optimised to simulate large IFUs. For example, a simulation of a single-pixel observation would take more than twice the actual observation time itself. Also, the user interface lacks flexibility as important features were hard-coded to fit the DESHIMA 2.0 instrument and optical path at the ASTE~\citep{Ezawa_2004} telescope. \texttt{maria} is an interesting option, as it is a general simulator for submm observations with an advanced atmospheric model~\citep{Morris2022}. However, it currently lacks direct support for spectroscopic detector configurations, also limiting its usability for simulating IFUs.

Therefore, we have developed \texttt{gateau} (Gpu-Accelerated Time-dEpendent observAtion simUlator), an end-to-end observation simulator for efficiently generating synthetic TODs of ground-based, single dish, astronomical observations with submm integral field units (IFUs) using KID technology. The simulator uses atmospheric screens to model the spatial/temporal atmosphere and a customisable radiative transfer cascade representing the optical path at the telescope to calculate the attenuation and contamination of the incoming astronomical source signal. The astronomical source is set by the user, according to their particular science case. For the dispersive element, \texttt{gateau} has a built-in filterbank model, but it also accepts user-supplied dispersive elements. This is useful, for example, when a fabricated filterbank response has been measured and a more accurate model is available, or when the dispersive element is not well described by a filterbank, which is the case for multichroic cameras. White noise is added to the output using a physically-motivated photon noise model. User-defined temporally correlated pink noise can also be included. \texttt{gateau} features a Python interface with a C/C++ backend, combining the ease of Python scripting with the raw computing power of compiled languages. Crucially, the \texttt{gateau} backend employs the CUDA programming model~\citep{Cheng2014} to achieve GPU-acceleration. This enables the simulation of large IFUs by exploiting the large number of computing threads present on GPUs~\citep{Kirk2017}, which allows for simulations to run orders of magnitude faster than the observation time. Also, \texttt{gateau} contains a random access memory (RAM) manager, which allows for simulations of any number of spaxel and voxel combinations whilst only being limited by the total available hard disk space.

The structure of the paper is as follows: in Sect.~\ref{sect:struct_imp} we give a general overview and describe the structure and software implementation of \texttt{gateau}. For each subsection, we discuss a particular component of \texttt{gateau}, how it is implemented, and what the main assumptions and limitations are for each implementation. In Sect.~\ref{sect:validations} we present comparisons of \texttt{gateau} simulations with actual DESHIMA 2.0 observations to validate the software. In Sect.~\ref{sect:example} we show an example use case for \texttt{gateau} by simulating an observation of a galaxy cluster through the Sunyaev-Zel'dovich effect~\citep{Sunyaev1970,Sunyaev1980}.

\section{Structure and implementation}
\label{sect:struct_imp}

\subsection{General overview}
\texttt{gateau} is an observation simulator for ground-based submm IFUs utilising KIDs. It models the radiative transfer of an arbitrary astronomical source signal through a dynamical model of the atmosphere and optical path at the telescope, down to the detector. The source can be spatially extended, and can have an arbitrary spectral shape. The radiative transfer cascade used between the atmosphere and detectors is set by the user and can consist of an arbitrary number of stages representing reflectors, dielectric lenses and cryostat windows, quasi-optical filters, and stray light baffles to name a few.

\begin{figure}[hbt]
\centering
    \includegraphics[width=0.4\textwidth]{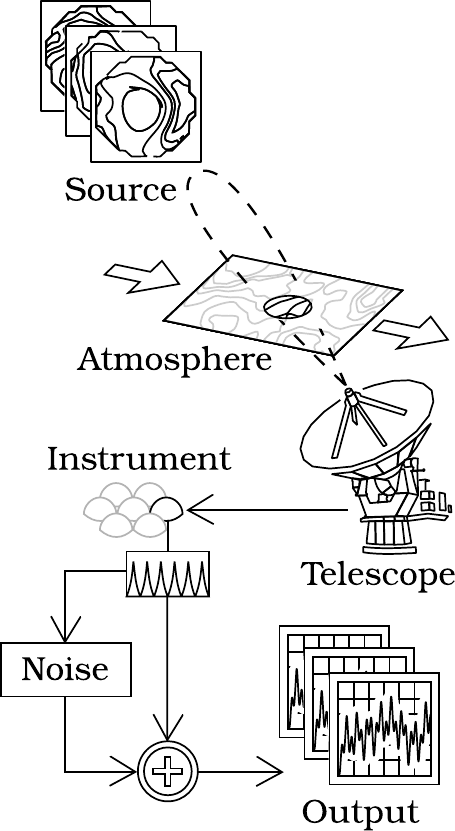}
    \caption{Graphical overview of a \texttt{gateau} simulation. The telescope scans the source and atmosphere screen. The source is depicted as multiple images, highlighting the spectral nature. The atmosphere screen is translated across the telescope beam with a constant windspeed, which is illustrated using the two arrows. After entering the telescope, the sky signal is propagated through the radiative transfer cascade and enters the instrument, illustrated here as a lenslet array representing individual spaxels. Here we highlight the signal entering a single spaxel, propagating through the dispersive element which we illustrate as a set of adjacent Lorentzian transmission curves. White and pink noise is then calculated and added to the received signal after the dispersive element. Finally, for each channel per spaxel, the output TODs are stored and the simulation is finished.}
    \label{fig:overview}
\end{figure}

The power received in the detector is used, together with a physical photon noise model, to calculate a white noise realisation which is added to the total power received. Additionally, pink noise can be added at the detector level to model, for example, readout or TLS noise. The output for each voxel in the IFU is a TOD in units of Watts, representing power entering the KID, or Kelvins, representing the brightness temperature seen. A graphical overview of a \texttt{gateau} simulation is given in Fig.~\ref{fig:overview}. A detailed, technical user guide can be found on the \texttt{gateau} documentation website\footnote{\url{https://tifuun.github.io/gateau/}}.

\subsection{Source and atmosphere}
\label{subsect:source_atm}
Before starting a \texttt{gateau} simulation, the astronomical source and atmosphere need to be specified and prepared. In \texttt{gateau}, an astronomical source cube is defined as an array of three dimensions, containing the specific intensity $I_\nu=I_\nu(\theta, \phi)$ as function of azimuth $\phi$, elevation $\theta$, and frequency $\nu$. The $\phi$ and $\theta$ coordinates are defined in a horizontal coordinate system centered on the telescope to be simulated and throughout the simulation the source remains stationary with respect to the telescope.

Then, $I_\nu$ is converted to a spectral power $P_{\nu,\mathrm{src}}$ through a 2-dimensional convolution by the telescope far-field pattern $F(\phi,\theta)$, and multipication by the physical telescope aperture area $A_\mathrm{p}$:
\begin{equation}
\label{eq:convolve_src}
    P_{\nu,\mathrm{src}}(\phi,\theta) = A_\mathrm{p}\iint I_\nu(\phi',\theta')F(\phi-\phi',\theta-\theta') \cos(\theta')\mathrm{d}\theta'\mathrm{d}\phi'.
\end{equation}
Note that $F(\phi,\theta)$ represents the gain pattern normalised by the peak amplitude, e.g. $\max(F(\phi,\theta))=1$. \texttt{gateau} provides a multi-threaded utility function for the calculation of $F(\phi,\theta)$ and the convolution in Eq.~\ref{eq:convolve_src}. This utility function requires the edge taper $t_\mathrm{edge}$ in decibels and the telescope aperture radius $R_\mathrm{tel}$. From this, $A_\mathrm{p}=\uppi R^2_\mathrm{tel}$ is calculated and the aperture illumination profile $A(l,m)$ is calculated assuming circularly symmetric Gaussian illumination at constant phase:
\begin{equation}
\label{eq:aper_dist}
    A(l,m) = 
    \begin{cases}
    \exp\left( -\frac{\lambda^2(l^2+m^2)}{R_\mathrm{tel}}2\log_{10}(-\frac{t_\mathrm{edge}}{20}) \right) & \text{if $\lambda^2(l^2+m^2)<R^2_\mathrm{tel}$},\\
    0 & \text{else},
    \end{cases}
\end{equation}
where $\lambda$ is the wavelength of light,  and $l$ and $m$ are the Cartesian aperture coordinates in units of $\lambda$. $F(\theta,\phi)$ is then obtained from $A(l,m)$ by taking the absolute value squared of the 2-dimensional Fourier transform:
\begin{equation}
    F(\phi,\theta) = |\mathscr{F}\{A(l,m)\}|^2,
\end{equation}
where $\mathscr{F}\{\cdot\}$ denotes the 2-dimensional Fourier transform. Setting the far-field by a direct Fourier transform of the aperture distribution relaxes the common assumption of a diffraction-limited beam and hence provides a more realistic situation. However, we do assume that the far-field pattern is independent of the telescope elevation. This is not generally true, as large aperture telescopes deform under gravity due to changing elevation angles~\citep{vonHoerner1975}. We also do not take into account error beams arising from correlated surface errors on the telescope primary~\citep{Bensch2001}, or diffraction due to support structures of the primary or secondary~\citep{Popping2007}. These assumptions reduce the input parameter set, whilst still providing a realistic model that captures the telescope main beam.

The atmosphere model currently used in \texttt{gateau} consists of user-supplied screens created by the Astronomical Radio Interferometer Simulator (ARIS)~\citep{Asaki2007} software, similarly to TiEMPO~\citep{Huijten2022}. ARIS creates a 2-dimensional screen of the extra path length (EPL) at zenith due to scattering in the atmosphere, assuming a Kolmogorov power spectrum of the spatial fluctuations in the atmospheric structure. Assuming that the EPL comes solely from fluctuations in precipitable water vapor (PWV) in the atmosphere, we can use the relation by~\cite{Smith1953} to convert the EPL screen to a PWV screen, assuming the EPL is solely due to water vapour and that the water vapour can be described as an ideal gas~\citep{Huijten2022}:
\begin{equation}
    \mathrm{PWV}(x,y) = \frac{\mathrm{EPL}(x,y)}{6.587} + \mathrm{PWV}_0,
\end{equation}
where PWV$_0$ is an average PWV value around which the fluctuations take place.

In order to convert the PWV screen to line-of-sight PWV, the screen is convolved by a 2-dimensional Gaussian with the footprint given in Eq.~\ref{eq:aper_dist}. Again, \texttt{gateau} provides a multi-threaded utility function for this convolution, which directly stores the output to disk in subscreens, so that the screens only have to be convolved once for each $R_\mathrm{tel}$ and $t_\mathrm{edge}$.

\subsection{Telescope and instrument}
\label{subsect:tel_inst}
After the source and atmosphere are defined, the telescope and instrument must be defined. For the telescope, parameters that can be set are the taper efficiency $\eta_\mathrm{t}$, a measure of how uniformly the antenna feed pattern illuminates the telescope aperture, and the root-mean-square (RMS) surface roughness $\rho_\mathrm{surf}$ from which the Ruze efficiency $\eta_\mathrm{surf}$ is calculated~\citep{Ruze1966}:
\begin{equation}
    \eta_\mathrm{surf} = \exp\left( - \left( \frac{4\uppi \rho_\mathrm{surf}}{\lambda} \right) \right)^2.
\end{equation}
The $\eta_\mathrm{surf}$ term, together with $\eta_\mathrm{t}$, will degrade the total source signal. This will be explained in more detail in Sect.~\ref{subsect:radtrans}.

The instrument is defined by setting the spectral and spatial characteristics, along with the material of the KID. Currently, for materials \texttt{gateau} only supports hybrid NbTiN-Al KIDs~\citep{Janssen2013,Reyes2026,Karatsu26}. The material choice sets quantities such as the pair breaking efficiency $\eta_\mathrm{pb}=0.4$~\citep{Guruswamy2014}, which quantifies what fraction of incoming power breaks Cooper pairs in the KID into quasiparticles, and the superconducting gap energy $\Delta_\mathrm{Al}=188$ $\upmu$eV. In turn, $\Delta_\mathrm{Al}$ also sets the low-frequency detection cutoff, i.e. photons with energy below $2\Delta_\mathrm{Al}$ are not detected. For the hybrid KIDs, this corresponds to a cutoff frequency of about 90 GHz. 

The spectral channels of the instrument are modeled by Lorentzians $\eta_\mathrm{filt}$, defined by a central frequency $f$ and resolving power $R$:
\begin{equation}
\label{eq:lorentzians}
    \eta_\mathrm{filt}(f,\nu) = \frac{\eta_\mathrm{peak}(f)}{4R^2(\nu/f-1)^2},
\end{equation}
where $\eta_\mathrm{peak}(f)$ is the peak height of each Lorentzian in isolation, i.e. when each spectral channel is fed individually by the instrument. With Eq.~\ref{eq:lorentzians}, it is directly possible to model dispersive elements such as gratings. However, to model a filterbank, some adjustments need to be made because filterbank channels are not individually fed, but connected to a common transmission line.

When placed in a filterbank it is customary to put the highest frequency filters at the beginning of the filterbank to minimise high-frequency losses in the transmission line~\citep{PascualLaguna2021}, and the lowest at the end. Because of this, low frequency filters will effectively see a signal from which the high frequency filters upstream have absorbed portions, decreasing the transmission in the downstream filters~\citep{marting26}. For a filter in the filterbank with some $f$, this can be modeled by subtracting the sum of $\eta_\mathrm{filt}(\bar{f})$ with $\bar{f} > f$:
\begin{equation}
\label{eq:filterbank}
    \eta_\mathrm{filt}(f,\nu) = \frac{\eta_\mathrm{peak}(f) - \sum_{\bar{f}>f}\eta_\mathrm{filt}(\bar{f},\nu)}{4R^2(\nu/f -1)^2}.
\end{equation}

The built-in filterbank model makes some major assumptions. For example, the model assumes that the power not transmitted through a filter to the KID it is coupled to travels downstream to lower frequency filters. In reality, part of the reflected power travels back upstream to higher frequency filters~\citep{PascualLaguna2021}, which requires modelling of a recursive redistribution of power across the filterbank. To properly model this would require significantly more information on the filterbank geometry, which we do not include in order to keep the input parameters minimal. Also, we ignore the coupling of the filter to harmonics above the fundamental resonance due to the difficulty of including these in a physically consistent way. However, in a well designed instrument these higher harmonics are suppressed and should not appear.

Aside from the filterbank or pure Lorentzian model, it is possible to directly supply the transmission curves to \texttt{gateau}. This is convenient when the dispersive element has been measured in the laboratory or obtained with a more advanced model than offered by \texttt{gateau}, increasing the accuracy of a simulation. Also, this option allows for the simulation of multichroic cameras, whose transmission curves are generally not well described by Lorentzians.

The built-in spaxel configuration is a packed hexagon, which can be specified by giving the circumradius of the hexagon $R_\mathrm{hex}$ in number of spaxels (excluding the central spaxel), and the far-field spaxel spacing $\delta_\mathrm{hex}$ along the circumradius. From these two numbers, \texttt{gateau} generates the far-field pointing offsets $(\Delta\phi_i,\Delta\theta_i)$ for each spaxel $i$ with respect to telescope broadside. It is also possible to supply custom pointing offsets per spaxel. 

\subsection{Scanning the source and atmosphere}
\label{subsect:scan}
Scan patterns in \texttt{gateau} are supplied during the initialisation of a simulation, and are performed in a horizontal coordinate system centered on the telescope. A scan pattern is a Python function mapping the simulation time $t$ onto a $(\phi(t),\theta(t))$ coordinate. It is performed around a constant azimuth and elevation offset $(\phi_0,\theta_0)$ which we will refer to as the scan offset. \texttt{gateau} contains two built-in scan patterns: a constant azimuth and elevation stare, and a simple daisy scan pattern based on the SCUBA-2 implementation~\citep{Scott05}:
\begin{align}
\begin{split}
\label{eq:daisy}
    \phi(t) &= r_\mathrm{petal}\cos(\Omega t)\sin(\omega t),\\
    \theta(t) &= r_\mathrm{petal}\sin(\Omega t)\sin(\omega t),
\end{split}
\end{align}
where $r_\mathrm{petal}$ is the radius of a petal of the daisy scan, $\Omega$ the angular frequency of a full daisy cycle, and $\omega$ the angular frequency of a single daisy petal. It is also possible to supply a user-defined scan pattern.

As the telescope scans the sky during a simulation, the telescope pointing $(\phi(t)+\phi_0,\theta(t)+\theta_0)$ is calculated by adding the scan pattern and scan offset. For each spaxel $i$, the spaxel pointing offsets $(\Delta\phi_i,\Delta\theta_i)$ are added to produce spaxel pointings $(\phi(t)+\phi_0+\Delta \phi_i,\theta(t)+\theta_0+\Delta \theta_i)$. For each spaxel pointing, a 2-dimensional linear interpolation is performed on $P_{\nu,\mathrm{src}}(\phi,\theta)$ to obtain the time-dependent source spectral power. If the spaxel pointing falls outside of the spatial grid of the source cube, the spectral power is set to zero.

At the same time, the atmosphere screen is translated with a constant windspeed $u_\mathrm{w}$ along the azimuthal direction. This assumes a "frozen-flow" atmospheric fluctuation model, which has been found to be a generally valid description using microwave and optical observations~\citep{Lay1997,Poyneer2009} and has been used, for example, in radio~\citep{Edler2021}. To convert the pointing of some spaxel $i$ at some time $t$ to an $(x_i(t),y_i(t))$ coordinate on the moving PWV screen, we assume a constant column height $h_\mathrm{column}$ above the telescope, within which we assume most of the PWV to be situated. Then, the atmospheric screen coordinates for spaxel $i$ are calculated using the scan pattern and spaxel pointing offsets:
\begin{align}
\begin{split}
\label{eq:xy_atm}
    x_i(t) &= h_\mathrm{column}\tan(\phi(t) + \Delta\phi_i ) + u_\mathrm{w}t,\\
    y_i(t) &= h_\mathrm{column}\tan(\theta(t) + \Delta\theta_i).
\end{split}
\end{align}
Using $(x_i(t),y_i(t))$, a 2-dimensional bilinear interpolation is performed on the atmosphere screen to obtain the PWV value corresponding to spaxel $i$ at time $t$. We then convert the PWV value to the frequency-dependent atmospheric transmission at zenith $\eta_\mathrm{atm,z}(t,\nu)$ using the Atmospheric Transmission at Microwaves (ATM)~\citep{Pardo2001} software, and correct for the elevation-dependent airmass~\citep{Guan2012}: 
\begin{equation}
\label{eq:airmass}
    \eta_{\mathrm{atm},i}(t,\nu)=\eta^{\mathrm{csc}(\theta(t)+\theta_0+\Delta \theta_i)}_\mathrm{atm,z}(t,\nu),
\end{equation}
where $\eta_{\mathrm{atm},i}$ denotes the airmass-corrected transmission for spaxel $i$. Together with specifying a physical temperature $T_\mathrm{atm}$ for the atmosphere, $\eta_\mathrm{atm}(t,\nu)$ fully describes the effect of the atmosphere on $P_{\nu,\mathrm{src}}(\phi,\theta)$ in \texttt{gateau}.

In order to increase computational efficiency, a 2-dimensional lookup table was created with ATM and used in simulations, instead of calculating $\eta_\mathrm{atm,z}$ in real-time. By using a lookup table for $\eta_\mathrm{atm,z}$, we are making the assumption that the only variables relevant in setting $\eta_\mathrm{atm,z}$ are $\nu$ and PWV. In reality, there are weak dependencies on the atmospheric pressure, humidity, and physical temperature of the atmosphere. Of these, the strongest is on temperature. However, the air temperature changes far more slowly than the PWV over an observation and therefore the time-dependent behaviour can be accurately captured by assuming that only the PWV affects $\eta_\mathrm{atm,z}$.

\subsection{Radiative transfer cascade}
\label{subsect:radtrans}
As the signal passes from the source to the instrument, it will encounter the atmosphere and several optical components, represented by stages in the radiative transfer cascade. Consider some spaxel $i$ at some time $t$. Let $N_\mathrm{stage}$ denote the total number of stages in the radiative transfer cascade, and let $k \in \{0,...,N_\mathrm{stage}-1 \}$ denote the index of a stage in the cascade. We explicitly include the initial pass of the source signal through the atmosphere, which will always be stage $k=0$, but exclude the final pass of the signal through the filterbank in $N_\mathrm{stage}$. We first convert the source $P_{\nu,\mathrm{src}}$, which has been interpolated on the spaxel pointing, into an effective $P_{\nu,0}$ seen by the 0th stage, by multiplying with the taper, Ruze, and polarisation efficiency terms~\citep{Gsten2006}, and adding the cosmic microwave background (CMB):
\begin{equation}
    P_{\nu,0}=\eta_\mathrm{t}\eta_\mathrm{surf}\eta_\mathrm{pol} P_{\nu,\mathrm{src}} + P^{\mathrm{JN}}_{\nu}(T_\mathrm{CMB}),
\end{equation}
where $\eta_\mathrm{pol}$ is the polarisation efficiency which is equal to 0.5 for a detector sensitive to a single polarisation and $P^{\mathrm{JN}}_{\nu}(T_\mathrm{CMB})$ the single-moded Johnson-Nyquist~\citep{Nyquist1928} spectral power due to a blackbody with the CMB monopole temperature $T_\mathrm{CMB}=2.725$ K~\citep{Fixsen2009}. 

Each stage $k$ modifies the incoming $P_{\nu,k}$, depending on whether the stage is reflective or refractive. A reflective stage is defined by a (possibly frequency-dependent) transmission efficiency term $\eta_k$, describing what fraction of the incoming spectral power is transmitted towards the next stage, and a single physical temperature $T_k$, which sets the parasitic Johnson-Nyquist spectral power mixed with the incoming signal. The spectral power after a reflective stage is given by:
\begin{equation}
\label{eq:radtrans}
    P_{\nu,k+1}=\eta_k(\nu) P_{\nu,k} + \left(1 - \eta_k(\nu) \right) P^{\mathrm{JN}}_{\nu}(T_k),
\end{equation}
where $\eta_k$ is the coupling efficiency of stage $k$. As mentioned before, $k=0$ is always the atmosphere stage, so that $\eta_0(\nu)=\eta_\mathrm{atm}(\nu)$ and $T_0=T_\mathrm{atm}$.

Ohmic reflector losses can be added to the cascade using the mechanisms of a reflective stage. The efficiency of an Ohmic loss stage is given by~\citep{Hagen1903}:
\begin{equation}
    \eta_k(\nu) = 1 - \epsilon_\mathrm{ref}\sqrt{\frac{\nu}{\nu_\mathrm{ref}}},
\end{equation}
where $\epsilon_\mathrm{ref}$ is a reference emissivity, measured at a reference frequency $\nu_\mathrm{ref}$, which we take to be $\epsilon_\mathrm{ref}=0.25\%$ at $\nu_\mathrm{ref}=840$ GHz for an aluminum reflector~\citep{Shitov08}. The frequency scaling of the reference value is valid for frequencies up to 30 THz~\citep{Silveira2010}, rendering it useful for \texttt{gateau}. 

In addition to having the stage parasitically couple to a temperature $T_k$, it is possible to let a stage couple parasitically to the atmosphere, in order to model losses to sky such as occurs with secondary spillover. In this case, $P^\mathrm{JN}_{\nu}(T_k)=(1-\eta_\mathrm{atm}(\nu))P^\mathrm{JN}_{\nu}(T_\mathrm{atm}) + \eta_\mathrm{atm}(\nu)P^\mathrm{JN}_{\nu}(T_\mathrm{CMB})$ and the calculation proceeds as usual.

In a refractive stage the transmission efficiency is calculated from the refractive index $n$, loss tangent $\tan\delta$, and thickness $d$ of the dielectric:
\begin{equation}
\label{eq:eta_refr}
    \eta^\mathrm{refr}_k(\nu) = \exp\left(\frac{-2 \uppi d n \tan\delta}{\lambda}\right).
\end{equation}
In addition to refraction, light transmitting through a refractive stage also undergoes a reflection at the dielectric surface upon entering the dielectric, and a reflection upon exiting, again at the surface. These are essentially reflective stages, for which the transmission efficiency is given by:
\begin{equation}
\label{eq:eta_refl}
    \eta^\mathrm{refl}_k = 1 - \left( \frac{1 - n}{1 + n}\right)^2,
\end{equation}
assuming a vacuum refractive index for the ambient space, perpendicular incidence of the light onto the dielectric surface, and that the light is either reflected or transmitted and not absorbed by the dielectric. Then, a full pass through a refractive stage is accomplished by applying Eq.~\ref{eq:radtrans} once with $\eta_k$ given by Eq.~\ref{eq:eta_refl}, then with $\eta_k$ given by Eq.~\ref{eq:eta_refr}, and finally once again with $\eta_k$ given by Eq.~\ref{eq:eta_refl}. Note that this only holds true when $d>\lambda$, which is typically the case for dielectric lenses or windows in the submm range. For the reflective and refractive parts of the stage, it is possible to supply different $T_k$, in case the dielectric temperature is different from the ambient temperature. It is also possible to omit the reflective parts of the refractive stage, mimicking the effect of anti-reflective (AR) coating.

The final pass of the signal through the filterbank is calculated in the following way:
\begin{equation}
    P_\mathrm{KID}(f)=\int \eta_\mathrm{filt}(f,\nu)P_{\nu} \mathrm{d}\nu,
\end{equation}
where $P_\mathrm{KID}$ denotes the total optical power entering the KID coupled to the filter at frequency $f$. Here, $P_\nu$ implicitly refers to the spectral power after the last stage in the cascade, i.e. $P_\nu=P_{\nu,N_\mathrm{stage}-1}$. We omit the $N_\mathrm{stage}-1$ subscript for clarity throughout the rest of the paper.

\subsection{Noise calculation}
The time-dependent $P_\nu(t)$ introduces a heteroscedastic Gaussian white noise component known as photon noise. This noise is modeled in \texttt{gateau} by three separate noise terms: shot noise, bunching noise, and generation-recombination noise. The noise terms are described by the expression for noise equivalent power (NEP)~\citep{Endo2019}:
\begin{equation}
    \mathrm{NEP}^2(t,f) = 2 \int \eta_\mathrm{filt}(f,\nu) P_{\nu}(t)\left( h\nu + \eta_\mathrm{filt}(f,\nu) P_{\nu}(t) + \frac{2\Delta_\mathrm{Al}}{\eta_\mathrm{pb}}\right) \mathrm{d} \nu,
\end{equation}
where $h$ is the Planck constant.

The calculated $\mathrm{NEP}$ is converted to standard deviation using the same method as in~\citet{Huijten2022}:
\begin{equation}
    \sigma(t,f)=\mathrm{NEP}(t,f)\sqrt{\frac{1}{2}f_\mathrm{sample}},
\end{equation}
where $f_\mathrm{sample}$ is the sampling frequency of the analogue-to-digital converter of the KID readout. Then, the power fluctuation TOD $P_\mathrm{wn}(t,f)$ due to white noise is drawn from a zero-mean normal distribution with standard deviation $\sigma(t,f)$:
\begin{equation}
    P_\mathrm{wn}(t,f)=\mathcal{N}(0,\sigma(t,f)).
\end{equation}

It is also possible to add stationary, temporally correlated noise, also known as pink noise. Throughout this paper, we will use the term pink noise to refer to any fluctuation with a power spectral density $\mathcal{P}_{xx}$ given by:
\begin{equation}
    \mathcal{P}_{xx}(f_{xx})\propto Af^{-\alpha}_{xx},
\end{equation}
where $f_{xx}$ is the noise frequency, defined as the frequency of fluctuations in the TOD, $\alpha$ is the pink noise slope, and $A$ is a scaling factor that sets the level of the pink noise. We then use the prescription described by~\citet{TimmerKonig95} to generate a power fluctuation TOD $P_\mathrm{pn}(t,f)$ due to pink noise.

\subsection{Simulation output}
\label{subsect:output}
The total power TOD $P^\mathrm{tot}_\mathrm{KID}(t,f)$ detected by a KID is given by the sum of the received power TODs from the radiative transfer cascade and the noise terms:
\begin{equation}
    P^\mathrm{tot}_\mathrm{KID}(t,f) = P_\mathrm{KID}(t,f) + P_\mathrm{wn}(t,f) + P_\mathrm{pn}(t,f).
\end{equation}
The output units of a simulation can be set to detected power or brightness temperature. For the latter output mode, \texttt{gateau} internally calculates the corresponding brightness scale by simulating a skydip observation~\citep{Archibald2002,Takekoshi2020}. The skydip method involves smoothly varying the telescope elevation over a specified range, smoothly varying the airmass, and hence the line-of-sight brightness temperature $T_b$. At the same time, the detector response is monitored and a mapping can be set up between the detector response and $T_b$. \texttt{gateau} accomplishes this by varying the actual PWV to vary $T_b$, and recording $P^\mathrm{tot}_\mathrm{KID}$ at the same time. Then a linear mapping is set up between $P^\mathrm{tot}_\mathrm{KID}$ and $T_b$. 

The output TODs of a simulation are directly stored to disk in HDF5 format~\citep{The_HDF_Group_Hierarchical_Data_Format}. The output file also contains extra output such as the zenith PWV, observation timestamps, and the telescope aperture efficiency. These can be used for the data reduction, if desired.

\section{Validations}
\label{sect:validations}
\subsection{DESHIMA 2.0 at ASTE model}
\label{subsect:deshima_model}
We validated \texttt{gateau} against observations obtained with DESHIMA 2.0 at the ASTE telescope~\citep{Ezawa_2004} in Chile. DESHIMA 2.0 is a wideband on-chip filterbank spectrometer operating between 200--400 GHz, using KIDs as the detecting element. ASTE is a 10-meter Cassegrain submm telescope, located at Pampa la Bola, Chile. During 2023--2024 DESHIMA 2.0 underwent its commissioning and science verification (CSV) campaign at ASTE and in order to faithfully compare the DESHIMA 2.0 observations to the \texttt{gateau} simulation, it is important to accurately model the setup at ASTE.

We extracted the taper efficiency $\eta_\mathrm{t}$ and spillover efficiency on the secondary from the laboratory measurements and simulations performed by~\citet{Moerman2024}, where they measured the complex beam pattern and propagated this through a simulated ASTE setup using the optical simulation software \texttt{PyPO}~\citep{Moerman2023}. We set $\eta_\mathrm{t}=0.85$. The edge taper of $-$10 dB was also obtained from these simulations. The ASTE RMS surface roughness is $42$ $\upmu$m (S. Ishii, priv. comm. 2016). The efficiencies of the optics between the secondary and cryostat were taken from design values. Cryostat window parameters were taken from~\citet{Takaku26}. Because the cryostat window is equipped with an AR coating, no reflections at the window interfaces were considered and we only supply the parasitic temperature of the window itself. The efficiency of the quasi-optical filter stack in the cryostat, cryostat optics, filterbank $\eta_\mathrm{peak}$, along with the filter $R$, central filter frequencies $f$, pink noise slope $\alpha$, and pink noise levels $A$, were taken from~\citet{Karatsu26}. The sampling frequency of the readout was set equal to the DESHIMA 2.0 sampling frequency $f_\mathrm{sample}=2\cdot10^9/2^{19}/24 \sim 159$ Hz. For the atmosphere, we set $h_\mathrm{column}=1537$ m~\citep{Corts2020}. We set the physical temperature of the atmosphere to 273 K. The values and physical temperatures used for the radiative transfer cascade can be found in Table~\ref{table:1}.

For all validation tests, we restricted the range of comparison to channels above 230 GHz. Between 200 and 230 GHz, a mismatch between the filterbank and the low-pass filterstack cutoff frequency of 460 GHz~\citep{Karatsu26} caused extra loading through the second harmonics of these filters. This is not modeled by \texttt{gateau}. We also only compare channels that were not flagged by the flagging procedure of DESHIMA 2.0.

\begin{table*}[ht!]
\caption{Efficiencies and physical temperatures used in the radiative transfer cascade for the validation tests. For the cryostat window, we give the window thickness $d$, refractive index $n$, and loss tangent $\tan\delta$. Since we disabled reflections at the window interface, we only supply the temperature of the window, and not the parasitic temperature of the surroundings seen in reflection.}                 
\label{table:1}    
\centering                        
\begin{tabular}{c c c c c c c}      
\hline\hline               
Stage description & $\eta_k$ & $d$ [mm] & $n$ & $\tan\delta$ & $T_k$ [K] & Notes\\         
\hline                      
   Primary spillover & 0.99 & - & - & - & 273 & Dabironezare et al. (in prep.) \\
   Primary Ohmic losses & "Ohmic-Al" & - & - & - & 273 & - \\
   Secondary Ohmic losses & "Ohmic-Al" & - & - & - & 273 & - \\
   Secondary spillover & 0.85 & - & - & - & "atmosphere" & \citet{Moerman2024} \\
   Warm optics spillover & 0.99 & - & - & - & 290 & \citet{Bosma21} \\
   Warm optics Ohmic losses & "Ohmic" & - & - & - & 290 & - \\
   Cryostat window & - & 4 & 3.41 & $10^{-4}$ & 290 & \citet{Takaku26} \\
   Low-pass filterstack & Custom & - & - & - & 4 & \citet{Karatsu26} \\
   Cold optics spillover & 0.95 & - & - & - & 4 & Dabironezare et al. (in prep.) \\
   Cold optics Ohmic losses & "Ohmic" & - & - & - & 4 & - \\
   Leaky lens efficiency & 0.7 & - & - & - & 4 & Dabironezare et al. (in prep.) \\
   peak filter efficiency & 0.16 & - & - & - & 0.12 & \citet{Karatsu26} \\
\hline                                  
\end{tabular}
\end{table*}

\subsection{Atmosphere observation}
\label{subsect:atm_val}
\begin{figure}
\centering
    \includegraphics[width=0.5\textwidth]{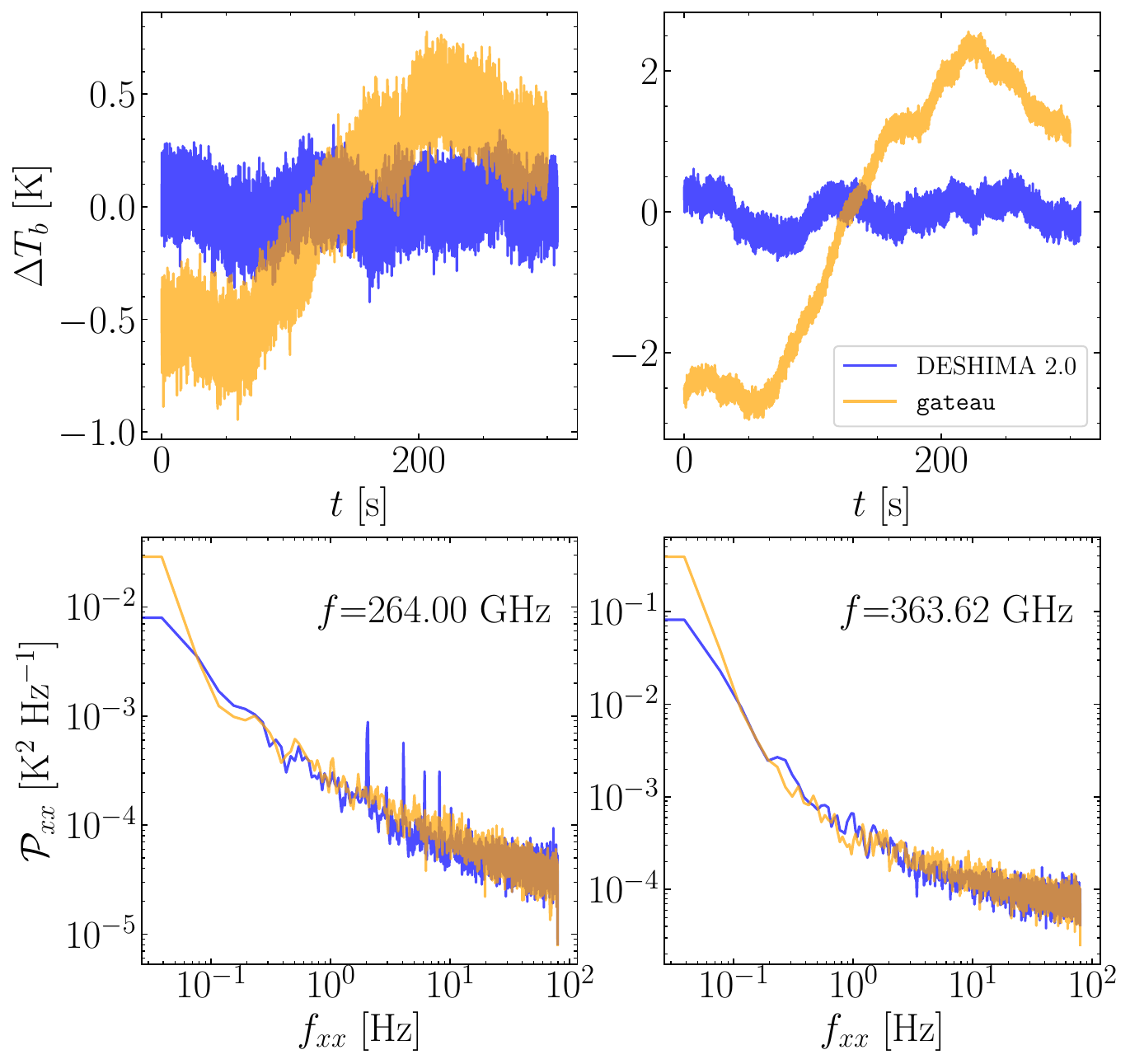}
    \caption{Example TODs and $\mathcal{P}_{xx}$ for two channels, about 100 GHz apart, obtained with DESHIMA 2.0 and simulated with \texttt{gateau}. Top row: average-subtracted brightness temperature $\Delta T_b$, as function of time. Bottom row: $\mathcal{P}_{xx}$ of the time series in the top row. Left column: channel at 250.54 GHz. Right column: channel at 349.18 GHz.}
    \label{fig:psd_tod_examples}
\end{figure}

For the first validation test we compare a \texttt{gateau} simulation of a blank atmosphere observation to a DESHIMA 2.0 blank atmosphere observation by comparing the temporal and spectral behaviour. The DESHIMA 2.0 observation was taken on October 20th at around 23:06 UTC, at a constant azimuth $\phi=180^\circ$ and elevation $\theta=60^\circ$, and took about 300 seconds. The measured maximum wind speed at the ASTE site was 14.77 m/s and the APEX radiometer reported a zenith PWV of about 0.9 mm. We applied these values to the atmosphere model. The DESHIMA 2.0 data was scaled to a brightness temperature scale $T_b$ using the skydip method described in~\citet{Takekoshi2020}. For both the DESHIMA 2.0 and \texttt{gateau} TODs, we calculated $\mathcal{P}_{xx}$ for each channel using the Welch method~\citep{Welch1967}. See Fig.\ref{fig:psd_tod_examples} for two examples.

In Fig.~\ref{fig:psd_tod_examples}, it can be seen that \texttt{gateau} shows more power for low $f_{xx}$. This can be attributed to the ARIS screen having more variation at longer timescales than the observed atmosphere, which is supported by the accompanying TODs in the top row.

\begin{figure}
\centering
    \includegraphics[width=0.5\textwidth]{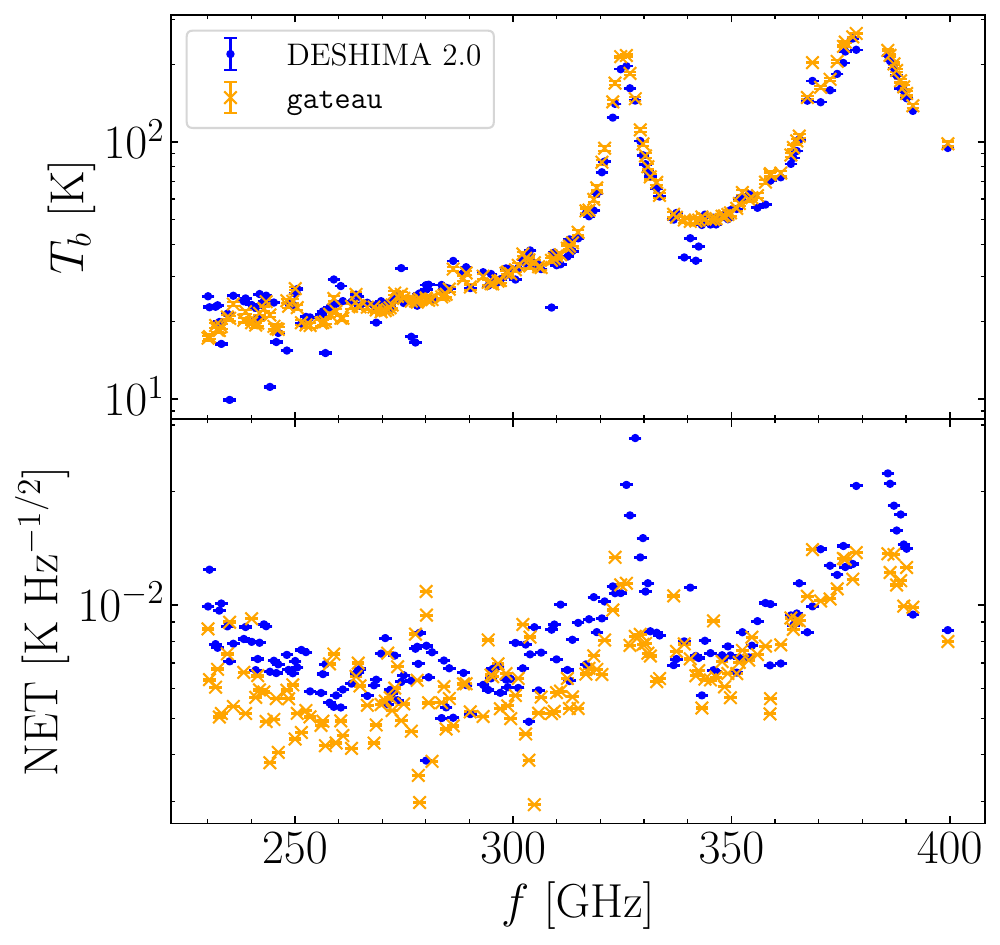}
    \caption{Comparison of atmosphere from DESHIMA 2.0 observation with \texttt{gateau} simulation. Top panel: spectra obtained by averaging the full observation. Bottom panel: comparison of per-channel NET, obtained from the average $\mathcal{P}_{xx}(f_{xx} > \text{10 Hz})$. The errorbars in both panels are at the 1-$\sigma$ standard error level.}
    \label{fig:validate_atm_net}
\end{figure}

For each channel, we extracted the noise-equivalent temperature (NET) by calculating the square root of the average $\mathcal{P}_{xx}$ for frequencies $f_{xx}$ higher than the crossover frequency for the atmospheric fluctuations and TLS noise, which is around 10 Hz for DESHIMA 2.0~\citep{Taniguchi_2022}. We also calculated the time-averaged atmospheric spectrum. The atmospheric spectrum and NET comparisons can be found in Fig.~\ref{fig:validate_atm_net}.

From Fig.~\ref{fig:validate_atm_net} we can see that the atmospheric spectrum is generally reproduced by \texttt{gateau}. There is larger scatter in the observed DESHIMA 2.0 spectrum, which could be due to errors introduced by the skydip method. The NET broadly agrees as well. However, the \texttt{gateau} NET seems to be slightly below the observed NET across the entire frequency range. This could be due to the actual ASTE PWV being slightly higher than the APEX PWV we used in the simulation, or the DESHIMA 2.0 filter profiles being slightly different from the Lorentzian filterbank model described in Eq.~\ref{eq:filterbank}.

\begin{figure}
\centering
    \includegraphics[width=0.5\textwidth]{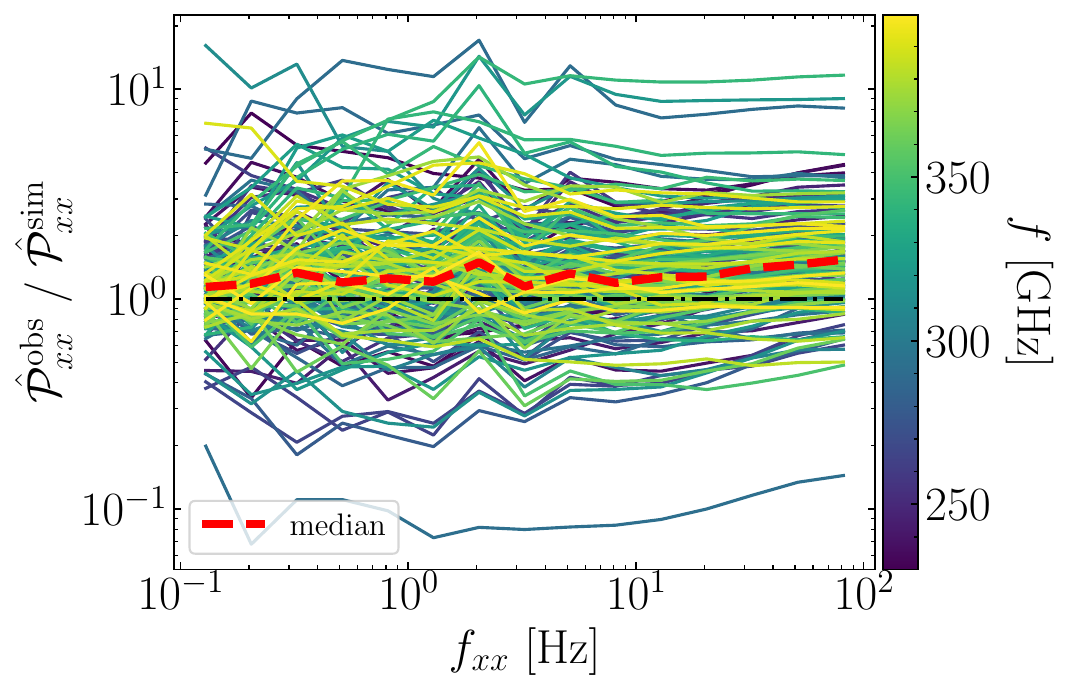}
    \caption{Per-channel power spectral density ratio $\hat{\mathcal{P}}^\mathrm{meas}_{xx}/\hat{\mathcal{P}}^\mathrm{sim}_{xx}$, as function of rebinned $f_{xx}$. The average APEX PWV during the observation was 0.9 mm. The ratio for each channel is color-coded to highlight the independence of the scatter  of the ratio on channel frequency. The median per $f_{xx}$ bin, across all channels, is illustrated as a red dashed line. For visual purposes, the horizontal black dash-dotted line denotes a ratio of 1, i.e. perfect agreement between DESHIMA 2.0 and \texttt{gateau}.}
    \label{fig:validate_ratio_psd}
\end{figure}

Further comparison was performed by comparing the actual $\mathcal{P}_{xx}$ between DESHIMA 2.0 and \texttt{gateau}. We emphasise that throughout this section, $\mathcal{P}_{xx}$ refers to the total power spectral density. This includes atmospheric fluctuations, pink noise, photon noise, and, for DESHIMA 2.0, noise sources that are not modeled in \texttt{gateau}. We rebinned the $\mathcal{P}_{xx}$ for both DESHIMA 2.0 and \texttt{gateau} to a common grid with bin edges equally spaced in logarithmic space. The rebinned power spectral densities are denoted $\hat{\mathcal{P}}^\mathrm{obs}_{xx}$ for DESHIMA 2.0 and $\hat{\mathcal{P}}^\mathrm{sim}_{xx}$ for $\texttt{gateau}$. The ratio $\hat{\mathcal{P}}^\mathrm{obs}_{xx}/\hat{\mathcal{P}}^\mathrm{sim}_{xx}$ can then be used to compare DESHIMA 2.0 and \texttt{gateau}, where a ratio of 1 indicates similar power at that specific $f_{xx}$. This comparison can be found in Fig.~\ref{fig:validate_ratio_psd}.

From Fig.~\ref{fig:validate_ratio_psd}, we see that, on a channel-to-channel comparison, the ratios vary substantially up to an order of magnitude. However, the median ratio across all channels shows a much closer agreement between DESHIMA 2.0 and \texttt{gateau}, to within a factor of 2. 

The discrepancy could be explained in several ways. Firstly, as was argued in Sect.~\ref{subsect:atm_val}, there could be a difference between the ASTE and APEX PWV or in the actual shapes of the filters. Secondly, different environmental and design factors affect the TLS level, such as cryogenic bath temperature~\citep{deGraaf2018,deRooij2021}, KID geometry~\citep{Gao2008-qh,Kouwenhoven2024}, readout tone frequency~\citep{Sueno2022}, and microwave readout power~\citep{Gao2007,Gao2008} in the KID. Thirdly, because the TLS noise depends on microwave power, there is also a dependence of the TLS level on optical loading. The optical loading shifts the KID resonance frequency, which decreases the absorbed microwave power at a fixed readout tone. In \texttt{gateau}, the dependency of TLS noise on optical loading is not simulated. Lastly, the simplified instrument model described in Sect.~\ref{subsect:tel_inst} could itself be a source of discrepancy.

Additionally, we also performed the same ratio comparison for two other observations, one with an APEX PWV of 0.6 mm and one with an APEX PWV of 1.7 mm. These can be found in Appendix~\ref{appendix:extra_compare}. We find that the ratio is more consistent with unity for the lower PWV scenario, but still find that the median ratio is within a factor 3 for the high PWV scenario.

\subsection{Daisy scan of compact source}
\label{subsect:uranus_obs}
For the second validation test we performed a comparison between an observation of a compact source with DESHIMA 2.0 and a simulation of \texttt{gateau}. Specifically, we will compare the extracted object spectra and channel-averaged maps. As compact source, we chose Uranus, which was observed regularly during the DESHIMA 2.0 CSV campaign as a primary flux/telescope pointing calibrator. The Uranus map was obtained in the daisy scanning mode. The particular observation we used was obtained on November 21st at 03:28 UTC at an elevation of $\theta=47.8^\circ$ and took approximately 4 minutes. Again, we used the skydip method to obtain a $T_b$ scale. The APEX PWV was 0.6 mm, which we applied to the atmosphere model. We set $u_\mathrm{w}$ to 10 m/s, as we wanted to see how well the simulation compared to the observation if we do not have access to windspeed information.

\begin{figure*}
\centering
    \includegraphics[width=\textwidth]{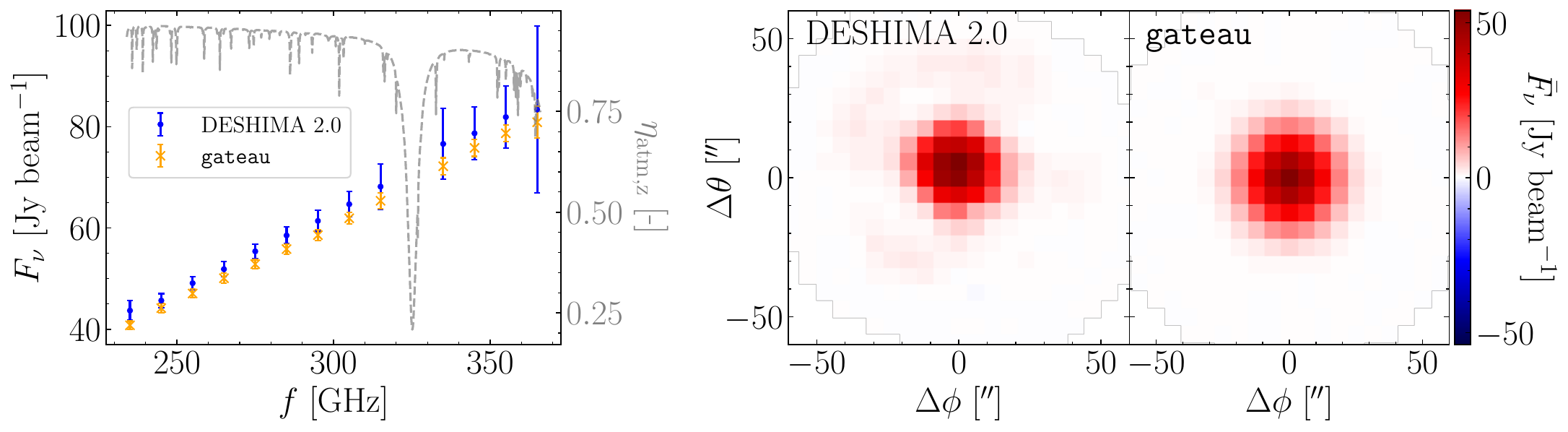}
    \caption{Comparison of a DESHIMA 2.0 observation of Uranus with a \texttt{gateau} simulation. Left panel: spectra extracted from the maps. The location in the maps was selected by picking the brightest pixel in the averaged flux density maps. The errorbars correspond to the 1-$\sigma$ standard error. We also show $\eta_\mathrm{atm,z}$ with PWV=0.6 mm as the grey dashed line to illustrate the atmospheric absorption peak and transmission windows. Right: channel-averaged flux density map for the DESHIMA 2.0 observation and \texttt{gateau} simulation.}
    \label{fig:validate_map_avg}
\end{figure*}

The time series outside of a $60^{\prime\prime}$ radius aperture were used, together with the calibrator wheel method~\citep{Ulich1976}, to convert $T_b$ to $T^\star_a$, the atmosphere-corrected antenna temperature. As flux calibrator we used a second Uranus observation, taken about 1 hour after the Uranus observation used for the validation, and also lasting about 4 minutes. The calibrator $T_b$ was converted to $T^\star_a$ in the same way as the validation observation. Both observations were then mapped onto a 2-dimensional grid with an extent of $\pm60^{\prime\prime}$ along both the azimuth and elevation axes, using an unweighted 2-dimensional binning. For both maps, we rebinned the channels to 13 new channels with a bandwidth of 10 GHz. We took care to exclude channels lying in or near frequency regions with strong atmospheric emission, most notably between 320-330 GHz and beyond 365 GHz. The rebinning was done in order to boost the S/N per channel map. This allowed us to fit a 2-dimensional Gaussian with high accuracy to each rebinned channel map, from which we extracted the peak $T^\star_a$ for both the validation and the calibrator Uranus map. Then, the flux conversion factor (FCF) was calculated for each rebinned channel using:
\begin{equation}
    \mathrm{FCF} = \frac{2k_\mathrm{B}T_{b,\mathrm{cal}}\Omega_\mathrm{cal}}{\lambda^2T^\star_{a,\mathrm{cal}}},
\end{equation}
where $k_\mathrm{B}$ is the Boltzmann constant, $T_{b,\mathrm{cal}}$ the model brightness temperature for the calibrator, $\Omega_\mathrm{cal}$ the solid angle subtended by the calibrator, and $T^\star_{a,\mathrm{cal}}$ the atmosphere-corrected antenna temperature of the observed calibrator in the rebinned channel. We used \texttt{radiobear}~\citep{dePater2014,dePater2019} to calculate the disk-averaged $T_b$ for Uranus, and the JPL ephemerides calculator~\citep{Giorgini1996} to retrieve $\Omega_\mathrm{cal}$.

The \texttt{gateau} model was made by starting with the disk-averaged $T_b(\nu)$ given by \texttt{radiobear}. After this, we defined a square spatial grid for the source, consisting of an azimuth-elevation distribution centered on $\phi_0=0^\circ$ and $\theta_0=47.8^\circ$, extending to $\pm70^{\prime\prime}$ in both directions, with 55 spatial grid points in both directions and 5000 source frequency points, uniformly distributed between 180 and 420 GHz. The frequency range of the source cube extends beyond the frequency range of the filterbank on purpose, in order to properly cover the transmission curves of the low and high frequency filters in the filterbank. Then, assuming the Rayleigh-Jeans law, we converted $T_b(\nu)$ to a specific intensity which we assigned to the central pixel in the grid:

\begin{equation}
I(\phi,\theta) = 
\begin{cases}
    C\frac{2 k_\mathrm{B} T_b}{\lambda^2} & \text{if $\theta=47.8^\circ$ and $\phi=0^\circ$},\\
    0 & \text{else},
\end{cases}
\end{equation}
where $C=\Omega_\mathrm{cal}/\Omega_\mathrm{pix}$ is a correction factor accounting for the fact that the source cube pixel solid angle and Uranus solid angle are different. Then, we apply the beam-source convolution described in Sect.~\ref{subsect:source_atm} to obtain the source cube used as input. We used the daisy scan as defined in Eq.~\ref{eq:daisy} with $r_\mathrm{petal}=65^{\prime\prime}$, $\omega=2\uppi\cdot1.54$ s$^{-1}$, and $\Omega=\omega / 50$. These values are representative of the values used at ASTE.

After simulating the observation, we applied the same reduction to the simulated TODs as with the observed ones. We also rebinned the channels in the same way. Again, we took the sections outside a $60^{\prime\prime}$ radius to subtract the atmosphere in order to convert $T_b$ to $T_a$, but instead of the chopper wheel method we took the average PWV over the observation to calculate the average $\eta_\mathrm{atm,z}$, and convert this to $\eta_\mathrm{atm}$ using Eq.~\ref{eq:airmass} and assuming a constant elevation $\theta_0$. Because we only have one spaxel in this experiment, we dropped the $i$ subscript from $\eta_\mathrm{atm}$. Finally, we divided $T_a$ by $\eta_\mathrm{atm}$ to obtain $T^\star_a$. Then, we used the aperture efficiency $\eta_\mathrm{ap}$, defined as $\eta_\mathrm{ap}\approx\eta_\mathrm{t}\eta_\mathrm{surf}\eta_\mathrm{so}$, with $\eta_\mathrm{so}$ being the spillover efficiency on the secondary reflector, and converted $T^\star_a$ to a flux density $F_\nu$ assuming the Rayleigh-Jeans law:
\begin{equation}
    F_\nu = \frac{2 k_\mathrm{B}T^\star_a}{A_\mathrm{p}\eta_\mathrm{ap}}.
\end{equation}

We then gridded the simulated TODs in the same way as the observed ones. In order to highlight systematics, the observed and simulated flux density maps were averaged over all channels using an inverse-variance weighting scheme to produce a spectrally-averaged flux density map $\bar{F}_\nu$. We selected the brightest pixel in these averaged maps, and extracted the spectra from the original rebinned maps using the brightest pixel location. We estimated the total standard error per channel by adding the standard error incurred by subtracting the off-source sections of the TODs and the standard error in the brightest pixel in quadrature. The brightest pixel spectra and averaged maps can be found in Fig.~\ref{fig:validate_map_avg}.

The spectra are consistent, but show a systematic discrepancy with \texttt{gateau} underestimating the flux density. Also, in regions where $\eta_\mathrm{atm,z}$ is smaller, the 1-$\sigma$ errorbars predicted by $\texttt{gateau}$ are significantly smaller compared to DESHIMA 2.0. The systematic discrepancies in the spectra in Fig.~\ref{fig:validate_map_avg} can be explained in several ways. First, the Uranus model in \texttt{gateau} is simplistic. For example, limb darkening has not been taken into account for the model. Also, strong polar brightening is present in resolved Uranus maps at submm wavelengths~\citep{Molter2021}, introducing spatial deviations in brightness temperature on the order of $\sim10$ K from the disk-averaged model we used. Moreover, the Uranian brightness temperature exhibits temporal variations~\citep{dePater2023} due to evolving structure and viewing angle, which is not modeled by \texttt{gateau}. The discrepancy in the 1-$\sigma$ error could come from the PWV during the DESHIMA 2.0 observation being slightly higher than the APEX PWV we used for the simulation, introducing a slighty larger photon noise level, or a higher level of fluctuations compared to the ARIS screens we used. These causes seem especially likely given that the discrepancy grows with decreasing $\eta_\mathrm{atm,z}$. Also, the difference in the scan patterns used could affect the total time spent on-source, affecting the error as well.

The spectrally-averaged maps also show general consistency, but \texttt{gateau} produces a slightly larger source in the map, indicating that the far-field beam used by \texttt{gateau} is larger than the DESHIMA 2.0 beam. This could be due to, for example, a higher edge taper for DESHIMA 2.0 at ASTE, which would make the far-field beam pattern smaller~\citep{Moerman2024,Karatsu26}. Additionally, the spectrally-averaged DESHIMA 2.0 map clearly shows elevated sidelobes. As mentioned in Sect.~\ref{subsect:tel_inst}, elevated sidelobe levels due to support structures of the telescope, or error beams due to correlated surface errors, are not taken into account, which could very well introduce the observed discrepancy. 

\section{Example use case}
\label{sect:example}
We conclude this work by presenting an example use case, where we simulate an observation of a faint source with an IFU model and attempt to make maps and spectra using the \texttt{gateau} output. For this example, we will use a simple mapmaking strategy involving a straightforward common-mode subtraction and 2-dimensional binning, rather than a more advanced mapmaking strategy. 

\subsection{IFU, telescope, and cascade models}
\label{subsect:ifu_tel_casc}
The IFU model was based on the successor of DESHIMA 2.0, the Terahertz Integrated Field Units with Universal Nanotechnology (TIFUUN) instrument. We set the channel range from 90 to 180 GHz and $R=20$, resulting in 15 channels per spaxel. The central frequencies $f$ were spaced such that each filter transmission curve overlapped adjacent curves at half-power points~\citep{PascualLaguna2021}. We set $R_\mathrm{hex}$ and $\delta_\mathrm{hex}$ to 3 spaxels and $102^{\prime\prime}$. This resulted in 37 spaxels, with a total of 555 KIDs. We added pink noise by randomly selecting 15 $A$ and $\alpha$ values from the full set mentioned in Sect.~\ref{subsect:deshima_model}.

The telescope model was the same as in Sect.~\ref{sect:validations}. We used the same daisy scan parameters as in Sect.~\ref{subsect:uranus_obs} and set the total observation time to 6 hours. For the atmosphere, we set $u_\mathrm{w}=10$ m/s and kept $h_\mathrm{column}$ and $T_\mathrm{atm}$ the same as in Sect.~\ref{subsect:deshima_model}. We set the average PWV to 0.8. The radiative transfer cascade is given in Table~\ref{table:2}.

\begin{table*}[ht!]
\caption{Efficiencies and physical temperatures used in the radiative transfer cascade for the example use case. For the lenses and cryostat window, we give $d$, $n$, and $\tan\delta$. We do not supply references in this table, because we do not intend to emulate an existing instrument and hence have freedom to choose what we deem appropriate.}                 
\label{table:2}    
\centering                        
\begin{tabular}{c c c c c c c}      
\hline\hline               
Stage description & $\eta_k$ & $d$ [mm] & $n$ & $\tan\delta$ & $T_k$ [K]\\         
\hline                      
   Primary spillover & 0.99 & - & - & - & 273\\
   Primary Ohmic losses & "Ohmic-Al" & - & - & - & 273\\
   Secondary Ohmic losses & "Ohmic-Al" & - & - & - & 273\\
   Secondary spillover & 0.85 & - & - & - & "atmosphere"\\
   Warm lens & - & 10 & 3.4 & $10^{-5}$ & 290\\
   Cryostat window & - & 4 & 3.4 & $10^{-5}$ & 290\\
   Low-pass filterstack & 0.65 & - & - & - & 4\\
   Cold lens 1 & - & 25.56 & 3.4 & $10^{-5}$ & 4\\
   Cold lens 2 & - & 16 & 3.4 & $10^{-5}$ & 4\\
   Leaky lens efficiency & 0.7 & - & - & - & 4\\
   peak filter efficiency & 1.0 & - & - & - & 0.1\\
\hline                                  
\end{tabular}
\end{table*}

\subsection{Source definition}
For the faint source, we generated a mock galaxy cluster, assuming an isothermal-$\beta$ model~\citep{Cavaliere1978}, and calculated the Sunyaev-Zel'dovich (SZ) signal~\citep{Sunyaev1970,Sunyaev1980}. We set the electron temperature $T_e=15.3$ keV, the central electron number density $n_0=10^{-2}$ cm$^{-3}$, peculiar bulk velocity $u_\mathrm{pec}=1000$ km/s with respect to Earth, cluster core radius $\theta_c=40.71^{\prime\prime}$, model parameter $\beta=0.73$, and we set the angular distance to $760$ Mpc. These parameters are based on parameters obtained from observations of CL 0016+16~\citep{Hughes1998,Worrall2003} except for the electron temperature, which was based on observations of RX J1347.5-1145~\citep{Kitayama2016}. We generated the model using \texttt{MockSZ}~\citep{mocksz}, a Python interface for generating mock SZ maps. The software follows the radiative transfer prescription given in~\citet{Birkinshaw1999} to calculate the relativistic thermal and kinematic SZ effect, by decoupling them and calculating them separately. In order to correct for the assumption of a decoupled thermal and kinematic effect, \texttt{MockSZ} adds relativistic cross-terms taken from~\citet{Itoh1998} and~\citet{Nozawa1998}. 

From the SZ model, we created a source cube defined on a square spatial grid, consisting of an azimuth-elevation distribution centered on $\phi=0^\circ$ and $\theta=47.8^\circ$, extending to $\pm210^{\prime\prime}$ in both directions, with 61 grid points along each axis. The source cube was convolved using Eq.~\ref{eq:convolve_src} before passing to \texttt{gateau}. The source cube contained 5000 frequency points, ranging from 89 to 260 GHz. We did not need to extend the low end far beyond the lowest frequency in the filterbank, because of the intrinsic cutoff at 90 GHz for the NbTiN-Al KIDs we simulated.

\subsection{Data reduction and results}
The zenith PWV included in the output was averaged over the entire observation and used to calculate $\eta_\mathrm{atm,z}(f)$. This was converted to $\eta_{\mathrm{atm},i}(t,f)$ using Eq.~\ref{eq:airmass}. Then, we calculated the $T^\star_{a,i}(t,f)$ for spaxel $i$ using common-mode subtraction in the following way:
\begin{equation}
\label{eq:common_mode}
    T^\star_{a,i}(t,f) = \frac{T_{b,i}(t,f)}{\eta_{\mathrm{atm},i}(t,f)} - \frac{1-\eta_{\mathrm{atm},i}(t,f)}{\eta_{\mathrm{atm},i}(t,f)N_\mathrm{out}}\sum_{j}^{N_\mathrm{spax}} \frac{S_jT_{b,j}(t,f)}{1-\eta_{\mathrm{atm},j}(t,f)},
\end{equation}
where $N_\mathrm{spax}$ is the total number of spaxels and $S_j$ is defined as follows:
\begin{equation}
\label{eq:Sj}
S_j = 
\begin{cases}
    1 & \text{if $\sqrt{\Delta\phi_j^2 + \Delta\theta_j^2} > r_\mathrm{out}$},\\
    0 & \text{else}.
\end{cases}
\end{equation}
Here, $r_\mathrm{out}$ is a radius specifying which spaxels are sufficiently off-source to be used for the common-mode average, which we set to $r_\mathrm{out}=200^{\prime\prime}$. In Eq.~\ref{eq:common_mode}, $N_\mathrm{out}$ is the number of spaxels satisfying Eq.~\ref{eq:Sj}, i.e. $N_\mathrm{out}=\sum_jS_j$.

\begin{figure*}
\centering
    \includegraphics[width=\textwidth]{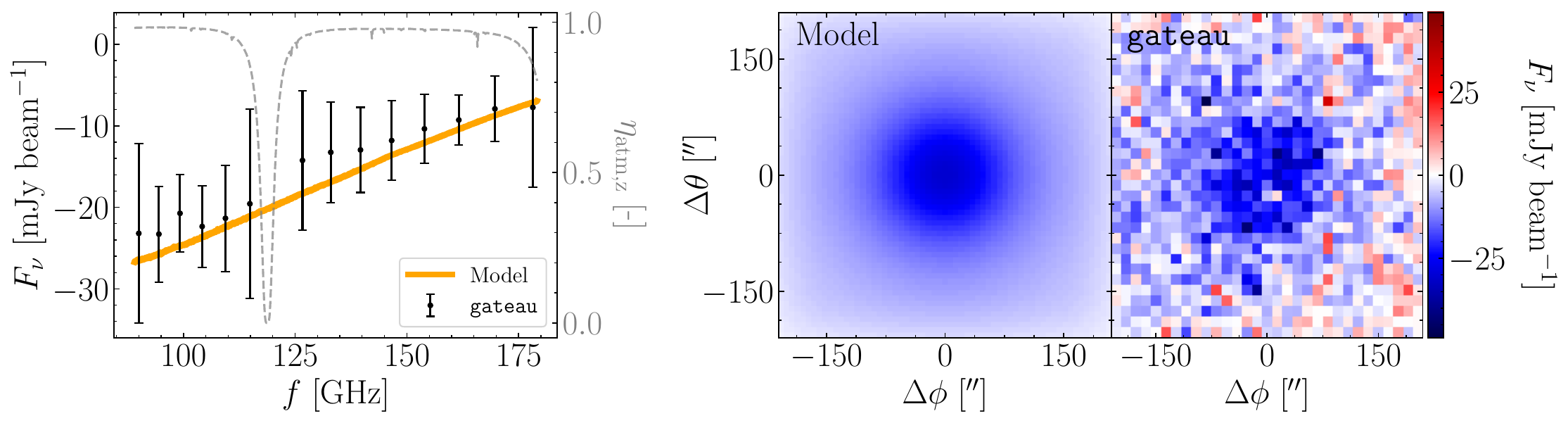}
    \caption{Overview of some results of the example use case. Left panel: averaged spectrum extracted from the created maps. The $F_\nu$ for each channel was obtained by averaging over a 60$^{\prime\prime}$ radius aperture centered on the cluster center. The errorbars are at 1-$\sigma$ standard error. For illustration, we also show the $\eta_\mathrm{atm,z}$ for PWV=0.8 mm as the dashed grey curve, in order to highlight the atmospheric absorption and transmission windows. Right panel: a comparison between the input model and the \texttt{gateau} output at 94.5 GHz. The input map shown here was averaged over a spectral range of 94.5$\pm 1$ GHz.}
    \label{fig:sz_use_case}
\end{figure*}

The TODs for each spaxel were then gridded onto a common 2-dimensional grid using the same method as Sect.~\ref{subsect:uranus_obs}, spanning $\pm210^{\prime\prime}$ with 31 gridpoints along both directions. In order to calculate the total value per pixel in the common grid, the contributions from each spaxel were averaged using an inverse-variance weighting scheme. An overview of some of the results can be found in Fig.~\ref{fig:sz_use_case}.

In Fig.~\ref{fig:sz_use_case} we show the spectrum in the left panel, obtained by averaging each channel map over a 60$^{\prime\prime}$ radius aperture, centered on $\phi_0$ and $\theta_0$. The standard error was calculated by adding the standard error per pixel inside the aperture and the standard error over the aperture average in quadrature. We do not include the errors incurred from common-mode subtraction and are hence underestimating the total error in the spectrum. We omit the channel around the absorption peak around 120 GHz, because the 1-$\sigma$ error was very large and dominated the figure. In orange, we show the input spectrum, obtained in the same way as the \texttt{gateau} spectrum, but applied to the input source cube. The right panel shows a comparison between the input source cube model and a \texttt{gateau} map at a channel frequency of 94.5 GHz, which we chose because it showed the strongest response out of all channels. The input model was averaged over a spectral range of 94.5$\pm 1$ GHz in order to create a map for comparison with \texttt{gateau}.

From Fig.~\ref{fig:sz_use_case}, we can see that the simulated maps resemble the input model in the center of the map. Towards the outskirts, the noise level increases as expected. Using such virtual observations, scan-patterns, calibration schemes and signal processing techniques can be tested against the model to optimize the experiment from end to end.

\subsection{Simulation-to-observation time ratio}
As a final test, we calculated the total per-spaxel simulation-to-observation time ratio $r_\mathrm{s/o}$ in order to give an intuition for the efficiency of \texttt{gateau}. For this, we used the same setup as in~\ref{subsect:ifu_tel_casc}, but we only simulated the central spaxel. We ran this test on a notebook equipped with an NVIDIA GeForce GTX 1650 Mobile (1024 CUDA cores and 4 GB RAM) and obtained a simulation time of 47 seconds, giving $r_\mathrm{s/o}\approx 0.0011$. For reference, we also ran the same experiment using TiEMPO, for which we found an $r_\mathrm{s/o}\approx2$ using 30 CPU threads, highlighting the efficiency of the GPU implementation of \texttt{gateau} compared to purely CPU-based implementations.

\section{Conclusion}
We presented \texttt{gateau}, a GPU-accelerated observation simulator for ground-based submm IFUs that use KIDs. The software is based on a Python interface, powered by a C/C++ backend that uses the CUDA API to achieve GPU-acceleration.

One of the main strengths, aside from its efficiency, is the modularity and flexibility of \texttt{gateau}. Because the source is user supplied, \texttt{gateau} can be used to simulate observations of a large range of astronomical sources and source fields. Also, one can set the telescope, atmosphere, and instrument parameters independently, allowing for multiple simulation scenarios for a single instrument. Moreover, the custom radiative transfer cascade allows for a flexible and realistic representation of the optical path at a telescope.

We validated \texttt{gateau} by comparison with DESHIMA 2.0 observations of the atmosphere and Uranus. From the comparison to the atmosphere, we find that \texttt{gateau} can qualitatively reproduce the photon noise and pink noise profiles of DESHIMA 2.0 at ASTE. The Uranus comparison showed that the observed and simulated spectra agree. Also, the averaged flux density maps are consistent. 

Furthermore, we performed a simulation of a realistic IFU model based on the TIFUUN design for the SZ science case, and spectrally mapped the output. We also show that the per-spaxel simulation time is orders of magnitude less than the observation time itself. We conclude that \texttt{gateau} is a suitable tool for simulating observations of different science cases for large KID-based submm IFUs. Therefore, it will prove valuable for the design and characterisation of such instruments, and the scientific interpretation of the obtained data.

Future work will involve adding the option of defining and scanning a source in equatorial coordinates, allowing for the testing of scanning strategies for very extended sources and source fields, such as for line intensity mapping. Another improvement will be to extend the atmosphere module to also work for atmosphere generators other than ARIS, which would extend the flexibility of \texttt{gateau} even further. An interesting option would even be to equip \texttt{gateau} with its own atmosphere generator, for example using the methods outlined by~\citet{Jia2015} or~\citet{Buscher2016}.

\section*{Software availability}
The \texttt{gateau} software, as well as the official documentation, is publicly available at \href{https://zenodo.org/records/17183878}{10.5281/zenodo.17183878}. The \texttt{MockSZ} software is publicly available at \href{https://zenodo.org/records/8365066}{10.5281/zenodo.8365066}.

\begin{acknowledgements}
This work is supported by the European Union (ERC Consolidator Grant No. 101043486 TIFUUN). Views and opinions expressed are however those of the authors only and do not necessarily reflect those of the European Union or the European Research Council Executive Agency. Neither the European Union nor the granting authority can be held responsible for them. A. M. thanks Michiel Darcis for discussions on the validity of using the frozen-flow atmospheric model. We thank Y. Asaki for kindly providing us with the ARIS software.

\end{acknowledgements}

\bibliographystyle{aa} 
\bibliography{refs}

\begin{appendix}
\section{Additional power spectra comparisons}
\label{appendix:extra_compare}
\begin{figure}
\centering
    \includegraphics[width=0.5\textwidth]{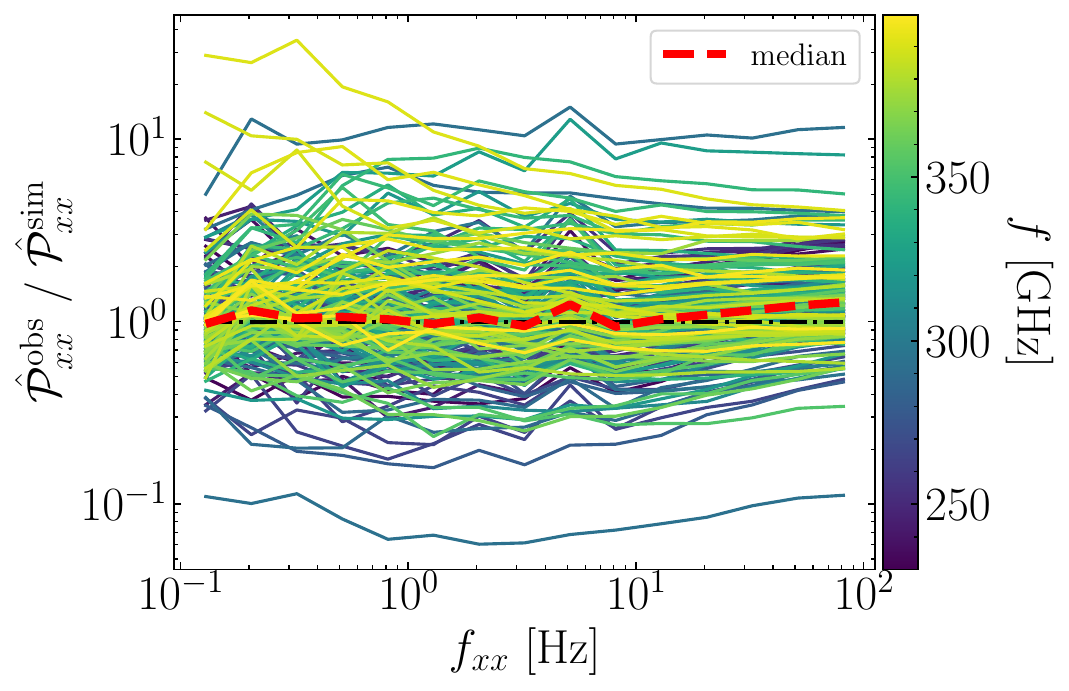}
    \caption{Ratios of $\hat{\mathcal{P}}_{xx}$ for the good weather scenario. The averaged APEX PWV was about 0.6 mm throughout the observation.}
    \label{fig:ratio_2}
\end{figure}

In addition to the $\hat{\mathcal{P}}_{xx}$ ratios shown in Fig.~\ref{fig:validate_ratio_psd}, here we show two other ratio comparisons. Both observations were 300 second stares of the atmosphere. They were chosen to represent different conditions compared to the one shown in Sect.~\ref{subsect:source_atm}. The first extra comparison, shown in Fig.~\ref{fig:ratio_2}, was chosen to represent a low PWV observation. The maximum windspeed was 15.14 m/s, and the APEX radiometer reported a PWV of about 0.6 mm.

\begin{figure}
\centering
    \includegraphics[width=0.5\textwidth]{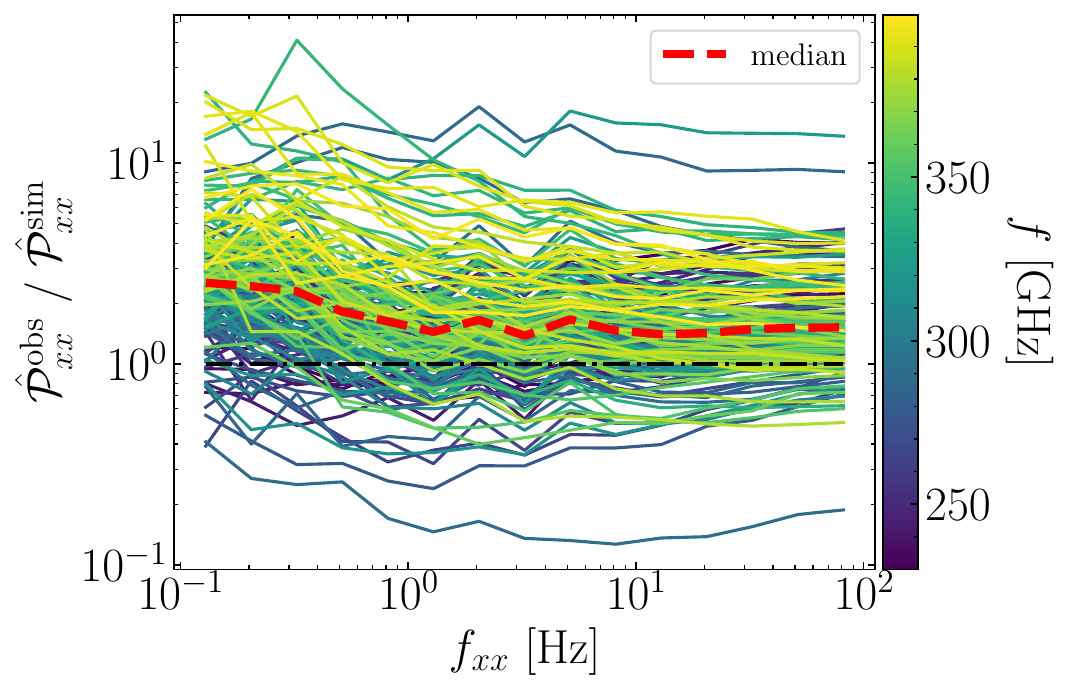}
    \caption{Ratios of $\hat{\mathcal{P}}_{xx}$ for the bad weather scenario. The averaged APEX PWV was about 1.7 mm throughout the observation.}
    \label{fig:ratio_3}
\end{figure}

As can be seen from Fig.~\ref{fig:ratio_2}, the median ratio is very close to one. The second extra comparison had a maximum windspeed of 14.3 m/s and the APEX radiometer reported a PWV of 1.7 mm. This ratio comparison can be found in Fig.~\ref{fig:ratio_3}.

\end{appendix}
\end{document}